\documentclass[12pt,preprint]{aastex}
\begin{document}

\title{Abundances in Turn-off Stars in the Old, Metal-Rich Open Cluster, NGC
6791} 

\author{Ann Merchant Boesgaard\altaffilmark{1}}
\affil{Institute for Astronomy, University of Hawai`i at M\-anoa, \\
2680 Woodlawn Drive, Honolulu, HI{\ \ } 96822 \\ }
\email{boes@ifa.hawaii.edu}

\author{Elizabeth E.~C.~Jensen}
\affil{Department of Mechanical and Aerospace Engineering, \\
Princeton University, Princeton, NJ{\ \ } 08544 \\ }
\email{ejtwo@princeton.edu}

\author{Constantine P.~Deliyannis\altaffilmark{1}}
\affil{Department of Astronomy, Indiana University\\
727 East 3rd Street, Swain Hall West 319, Bloomington, IN {\ \ }47405-7105}
\email{cpd@astro.indiana.edu}

\altaffiltext{1}{Visiting Astronomer, W.~M.~Keck Observatory jointly operated
by the California Institute of Technology and the University of California.}

\begin{abstract}
Open clusters have long been used to illuminate both stellar evolution and
Galactic evolution.  The oldest clusters, though rather rare, can reveal the
chemical and nucleosynthetic processes early in the history of the Galaxy.  We
have studied two turn-off stars in the old, metal-rich open cluster, NGC 6791.
The Keck + HIRES spectra have a resolution of $\sim$45,000 and signal-to-noise
ratios of $\sim$40 per pixel.  We confirm the high value for [Fe/H] finding
+0.30 $\pm$0.08, in agreement with earlier results from evolved stars in other
parts of the HR diagram.  We have also determined abundances for Na, Si, Ca,
Ti, Cr, Ni, Y and Ba.  These are compared to a sample of old, metal-rich field
stars.  With the probable exception of enhanced Ni in the cluster stars, the
field and cluster stars show similar abundances of the elements.  Model
predictions show that the Ni enhancement could result from enrichment of the
pre-cluster gas by SN Ia.  Orbital evidence indicates that NGC 6791 could have
originated near the inner regions of the Galaxy where the metallicity is
generally higher than it is in the disk or halo.  Subsequent perturbations and
migrations may have resulted in its current heliocentric distance of 4 kpc and
1 kpc above the Galactic plane.
\end{abstract}


\section{INTRODUCTION}

The study of open clusters has greatly advanced our understanding of stellar
evolution through comparisons between observations and theoretical models.
The distribution of stars in the HR diagram has been an especially useful tool
in this regard.  Comparing clusters of different ages via the turnoff points,
as shown initially by e.g. Johnson \& Sandage (1955), Sandage (1956), has
proven to be extremely important.  Open clusters have also been used to
increase our knowledge of cluster and Galactic dynamics, the distance scale of
the universe, and the chemical history of the Galaxy.

Stars within a particular cluster are thought to have been formed together,
within a few Myr, and share the same composition and space motion.
Comparative composition studies have mostly been limited to Fe and Li,
however.  See Friel et al.~(2002) for a presentation of metallicities and
kinematics for 39 old open clusters (older than 0.8 Gyr).  However, it is
important to study a full range of elements including CNO and other
alpha-elements, Fe-peak elements, n-capture elements.  Cluster-to-cluster
comparisons can reveal possible trends with age.  Comparisons between dwarfs
and giants reveal potential mixing that could alter the surface abundances in
the evolved giants.  Such comparisons have been done for globular clusters
providing interesting insights into their evolution, but open clusters have
not yet received such comprehensive studies.  Old clusters are especially
interestesting as they are nearly as old as the Galactic disk.

Ages can be determined for open clusters and they are found to span the
lifetime of the Galactic disk.  Open clusters that are old are relatively rare
because old clusters are subject to dissipation after several passages through
the disk plane such that former cluster members become part of the general
field.  The field stars of Chen et al.~(2003) and Feltzing \& Gonzales (2001)
that are old and metal-rich could be the remnants of similar, but less
massive, clusters like NGC 6791.  The extant old clusters tend to probe the
outer regions of the Galaxy and are good tracers of Galactic chemical
evolution.  Current evidence seems to point to a rapid, early chemical
enrichment of the disk, plus infall of some metal-poor gas (e.g. Friel et
al.~2002).  This picture comes from abundances of Fe only.  A much fuller
picture would require abundances for an array of elements since different mass
ranges of stars produce different nucleosynthetic products.

The old open clusters contain the clues to early chemical evolution in the
atmospheres of their unevolved stars.  Unlike many of the giant stars, their
atmospheres are not contaminated by the products of nuclear reactions in the
interiors.

\section{NGC 6791}

The cluster, NGC 6791, is an enigmatic open cluster.  It is unusually massive;
apparently it's metallicity is at least a factor of two higher than the sun's,
yet it is very old.  In addition, it has extreme kinematics and is about 1 kpc
above the Galactic plane.  This cluster defies many of the existing paradigms
about the formation of the Galaxy and its chemical history.  Thus it is
especially important to determine the details of its chemical composition.

NGC 6791 is anomalous in several respects: (1) Typical open clusters contain a
few hundred stars, but NGC 6791 is very populous with a mass of $\sim$4000
M$_{\odot}$ (Carraro et al.~2006, Gratton et al.~2006).  (2) It is apparently
super metal-rich with [Fe/H] $\sim$+0.4 -- but this has been determined only
from its evolved stars (Peterson \& Green 1998, Worthey \& Jowett 2003,
Gratton et al.~2006, Carraro et al.~2006).  (3) In spite of being metal-rich,
it is very old; the photometry is excellent (Montgomery et al.~1994) and the
HR diagrams yield ages of 8-10 Gyr (Demarque et al.~1992, Garnavich et
al.~1994, Montgomery et al.~1994, Chaboyer et al.~1999).  Other old open
clusters include M67 and NGC 188 which have solar metallicity and ages of 4-7
Gyr.  (4) It clearly does not fit any age-metallicity relation for the
Galactic disk given its very old age and unusually high metallicity
(e.g. Carraro et al.~2006).  (5) The combination of its heliocentric distance
of $\sim$4 kpc (King et al.~2005) and its Galactic latitude of +11$^{\circ}$,
means it is $\sim$1 kpc above the Galactic plane, far above the disk where
open clusters reside.  (6) Given its high metallicity and its distance from
the Galactic center (Bedin et al.~2006), it does not fit any radial abundance
gradient for the Galaxy (e.g Carraro et al.~2006).  (7) It has an atypical
white dwarf cooling curve (Bedin et al.~2005).

Photometry of NGC 6791 has been done by Kinman (1965), Kaluzny (1990),
Montgomery, Janes \& Phelps (1994) (hereafter MJP), Kaluzny \& Rucinski
(1995), Stetson et al.~(2003), Carney et al.~(2005) and Anthony-Twarog et
al.~(2007).  Isochrone fitting in the color-magnitude diagram indicates both
high metal content and an old age for NGC 6791.

The high metal content has been confirmed by spectroscopic studies.  The Fe
abundances that have been found so far for this cluster are from evolved
stars.  The first spectroscopic alert that the cluster showed high metallicity
came from Peterson \& Green (1998) who observed one blue horizontal branch
star at medium-high resolution (20,000) with the echelle spectrograph on the
NOAO/KPNO 4 m Mayall telescope.  They found a temperature of 7300 $\pm$ 50 K
and log g = 3.6 $\pm$ 0.2 dex resulting in an iron abundance of [Fe/H] = +0.4
$\pm$ 0.1 dex.  Based on the proper motion and radial velocity their star is a
confirmed member of the cluster.  Worthey \& Jowett (2003) looked at 24 giant
stars at medium-low resolution ($\sim$2,200) with the MDM 2.4 m telescope.
Their mean iron abundance is [Fe/H] = 0.320 $\pm$ 0.023 dex.  Gratton et
al.~(2006) observed four red giant clump stars at a resolution 29,000 with the
SARG spectrograph on the Galileo National Telescope.  They found [Fe/H] =
+0.47 ($\pm$ 0.04, rms = 0.08) dex.  Carraro et al.~(2006) used HYDRA on the
WIYN telescope to obtain echelle spectra of ten giant stars with a resolution
of 20,000 and found the metal abundance to be [M/H] = +0.39 $\pm$ 0.05.

Recent work from deep imaging with HST second epoch observations have allowed
the determination of the cluster's proper motion and orbit (Bedin et
al.~2006).  According to these authors this cluster had its origin in the
inner regions of the Galaxy.  It has come within 3 kpc of the Galactic center
and has crossed the disk in regions of high stellar density.  The cluster has
survived because of its high stellar mass and density.  The orbit has an
unusually high eccentricity for an open cluster, $\epsilon$ $\sim$0.5.  It
travels to an outer distance from the Galactic center of $\sim$10 kpc.

In this work we have determined abundances of several elements in two
``unevolved'' main-sequence turn-off stars.  Their surface composition is
unaltered by nuclear fusion products and subsequent mixing, which affects the
red giant and horizontal branch stars.  Our stars represent the original
composition of the stars and the cluster, with the potential exception of the
effects of microscopic diffusion.  Richard et al.~(2002) have studied the
effects of gravitational settling, thermal diffusion, and radiative
acceleration on the surface composition of metal poor stars with models with
[Fe/H] from $-$4.31 to $-$0.71; they have shown that the effects for globular
cluster turnoff stars are much smaller in the higher metallicity models.
Michaud et al.~(2004) suugest that turnoff stars may show small
underabundances relative to the initial abundances.  For the open cluster M67
at 3.7 Gyr they calculate that [Fe/H] may be down by $-$0.03 dex and [Ti/H]
down by $-$0.02 at 5800 K in turnoff stars.  For the open cluster NGC 188 with
an older age of 6.7 Gyr, they find that [Fe/H] may be down by $-$0.05 dex and
[Ti/H] by $-$0.04 dex.  Our cluster, NGC 6791, is 8-10 Gyr in age and would
have had more time for microscopic diffusion to have taken place.  Still, the
effects at 5800 K are small and even smaller at lower temperatures.

\section{OBSERVATIONS}

Figure 1 shows the HR diagram with the photometry of MJP where the positions
of the two stars we observed indicated by large circles.  We have obtained
high-resolution, high signal-to-noise echelle spectra from the Keck I 10-m
telescope with the original HIRES (Vogt et al.~1994) of those two turnoff
stars in NGC 6791: MJP 4112 and MJP 5061.  The turn-off stars are faint at
V$\sim$17.4, so long integrations were needed even on the largest telescope.
The data were obtained on two observing runs, 1999 June 7 and 8 and 2000 May
28 and 29 (UT). A log of the observations is presented in Table 1.  The
individual integration times were 40 to 60 min with a total of four hours on
MJP 4112 and three hours on MJP 5061.  The spectral resolution is $\sim$45,000
and the signal-to-noise ratio $\sim$40 per pix.  Exposures were made of 15
internal flat fields and 15 bias spectra; comparison spectra of Th-Ar were
obtained at the beginning and end of each of the four nights.

We conducted standard data reduction procedures in IRAF.  The images were
first bias-subtracted.  A master-flat field was created by median-combining
all of the flats and then normalized using the routine {\em flatnormalize}.
The images were then divided by the master flat and cosmic ray events were
removed.  The spectra, including the Th-Ar comparisons were traced and
extracted.  The wavelength solution from the Th-Ar spectra was applied to the
individual star spectra, and then the spectra were corrected for their radial
velocity shifts.  The multiple exposures for the individual stars were median
combined and the continuum level was found by fitting a high-order spline3
function to the individual orders.  Figure 2 shows a sample of the reduced,
combined spectra of each star near 6160 \AA.  Lines of Ca I, Si I, and Fe I
are indicated.  We took a 10 s exposure of the spectrum of the Moon, as a
solar surrogate, with the same equipment on 2003 Jan.~11; this is shown
for comparison in the middle panel of the figure.  The lines in both cluster
stars are clearly stronger than the solar lines.

We measured equivalent widths of nearly 140 lines in each star over several
orders, mostly between 5900 and 6800 \AA\ because there was considerable
blending at shorter wavelengths.  The measurements were made in IRAF with the
{\em splot} package with Gaussian profile fitting.  We measured lines of the
alkali element, Na (Na I), the $\alpha$-fusion elements, Si, Ca, Ti (Si I, Ca
I, Ti I), the Fe peak elements, Cr, Fe, Ni (Cr I, Cr II, Fe I, Fe II, Ni I)
and the neutron-capture elements, Y (Y II) and Ba (Ba II).  A table of these
line measurements, along with the species, wavelengths, lower excitation
potentials, and oscillator strengths ($log~gf$) values is available
in the Appendix as Table 2.

\section{ABUNDANCES}

\subsection{Stellar Parameters}

Due to the low galactic latitude, 11$^{\circ}$, and large distance, $\sim$4000
pc, of NGC 6791, reddening is a problem.  And due to the high metallicity of
NGC 6791 it is difficult to calibrate the reddening effects.  That results in
an uncertain reddening correction.  In fact, published values range from
$E(B-V)$ = 0.10 (Janes 1984) to 0.22 (Kinman 1965).  Liebert et al.~(1994)
found $E(B-V)$ = 0.135, Chaboyer et al.~(1999) concluded that the reddening is
between 0.08 and 0.13, while Anthony-Twarog (2007) found $E(B-V)$ = 0.155
$\pm$0.016 based on Str\"omgren photometry.  The latter authors point out that
the usual calibrations are likely to underestimate the colors of turnoff stars
with high metallicity.  The uncertainty in the reddening and in the
calibrations means that temperatures determined from color indices will be
uncertain.  Therefore, we did not try to determine temperatures from (B$-$V)
or other color indices.  We chose to find T$_{\rm eff}$ spectroscopically by
making use of the plentiful lines of Fe I.  Thus we avoided the problems with
the effect of interstellar reddening on the colors and thus the temperatures
derived from them.  We point out, however, that the model atmospheres will be
uncertain due to the high metallicity and this may introduce some systematic
errors.


For the temperature determinations we used $\sim$35 Fe I lines ranging in
excitation potential from 0.86 to 4.83 eV.  We tried to determine values for
log g from comparison of abundances found from Fe I and Fe II and from Cr I
and Cr II; however there were few usable lines of either Fe II and Cr II.  We
settled on log g of 4.3 which gave a good compromise for Fe and Cr in MJP 5061
for the agreement for the two ions in the two elements.  We used the same
value for log g for MJP 4112 based on the similarity of the positions of the
two stars in the color-magnitude diagram.  We calculated the microturbulent
velocity, $\xi$, from the empirical formula involving T$_{\rm eff}$ and log g
from Edvardsson et al.~(1993); this relationship is not validated in stars of
such high metallicity, but our results are not overly sensitive to the value
of the microturbulent velocity.  The stellar parameters we used for the model
atmospheres are given in Table 3.

\subsection{Elemental Abundances}

We employed MOOG (2002 version) to determine abundances of several elements
via the {\it abfind} driver.  The grid of model atmospheres of Kurucz was
interpolated to find the model atmosphere for each star.  We started with
models with [Fe/H] = +0.4, but it became apparent that the abundance of Fe was
closer to +0.3, so we rederived all the abundances with that value for [Fe/H]
in the models.  Abundances were determined for each ion.  For elements with
more than one ion, the final abundance was determined as weighted by the
number of lines used.  For Ba the hyperfine structure was included in the line
input data and the {\it blends} driver was used in MOOG.  The results can be
found in Table 4.  The standard deviations from the internal agreements from
individual lines are given there.

We were unable to derive an abundance for Li because the line at 6707.45,
presumably due to Fe I, was much stronger than the Li feature in the blend due
to the high value of [Fe/H].  An attempt at spectrum synthesis of the feature
was unsuccessful due to the weakness of the Li doublet and the noise in the
spectra; we were able to obtain a satisfactory fit to the 3 Fe I lines in the
region: $\lambda\lambda$6703, 6705, 6710 in both stars.

We determined abundance errors that result from the uncertainties in the model
parameters by running additional models which varied T$_{\rm eff}$ by 100 K,
varied log g by 0.2 dex, [Fe/H] by 0.1 dex and  $\xi$ by 0.2 km s$^{-1}$.
Tables 5 and 6 show those errors for each parameter for each ion in each
star.  

\section{RESULTS}

The abundance ratios that we derived are presented in Table 7.  The 1$\sigma$
errors given in that tables are the result of adding the errors in Tables 5
and 6 in quadrature.  (They do not include uncertainties in the equivalent
width measurements.)  The cluster mean, the average for the two stars, is also
given in Table 7 with the 1$\sigma$ error taken from the individual errors as
divided by the square root of 2. 

Our Fe abundances confirm the high Fe content found in the evolved stars in
NGC 6791 -- giants and a horizontal branch star -- from lower resolution
spectroscopy.  Based on 32 Fe I lines and five Fe II lines we find log N(Fe/H)
= 7.81 and [Fe/H] = 0.29 $\pm$ 0.11 for MJP 4112.  For MJP 5061 we used 36 Fe
I lines and four Fe II to find log N(Fe/H) = 7.83 and [Fe/H] = 0.31 $\pm$
0.11.  The average for the two turnoff stars is thus [Fe/H] = 0.30 $\pm$0.08.
(For solar log N(Fe/H) we adopt the value used in the MOOG program from Anders
and Grevesse (1989) of 7.52.)

In addition we have found elemental abundances for eight other elements.
Figure 3 shows the abundances in our two stars as normalized to Fe,
i.e. [X/Fe], plotted against their values of [Fe/H] for those eight elements.
The cluster mean error bars are shown in the upper right of each panel.
Within that error we see that the alkali metal, [Na/Fe], and the alpha
elements, [Si/Fe], [Ca/Fe], and [Ti/Fe] are near 0.0, i.e., normal.  The
neutron-capture elements, [Y/Fe] and [Ba/Fe] also appear to be normal.  That
is, those elements are enhanced as much as Fe is enhanced relative to solar.
The Fe-peak element, Cr, is also in alignment with Fe.  However, both stars
seem to have enhanced values of [Ni/Fe], another Fe-peak element.

We can compare our results with those of field star samples that are both old
and metal-rich.  In the Edvardsson et al.~(1993) sample of 189 disk stars
there are six stars that are older than 8.5 Gyr with [Fe/H] = +0.01 to +0.13.
In the spectral abundance analysis of old, metal-rich disk field stars of Chen
et al.~(2003) there are nine stars with ages $>$8 Gyr and with [Fe/H] values
from +0.05 to +0.47.  In addition, Feltzing \& Gonzales (2001) have four stars
that are older than 8 Gyr and a range in [Fe/H] from +0.15 to +0.47.  The
elements in common in our analysis and those studies are Na, Si, Ca, Ti, Cr,
and Ni, although Cr was not done by Edvardsson et al.~(1993).  This comparison
is shown in Figure 4.  Our NGC 6791 turnoff stars are remarkably similar to
those old, metal-rich field stars.  The ratio, [Na/Fe], is near 0.0 in NGC
6791 and in most of the field stars.  The alpha element ratios, [Si/Fe],
[Ca/Fe], and [Ti/Fe] are in agreement in the field and cluster stars.  The
Fe-peak element, Cr, should be near 0.0 and it is within the errors.  Compared
to those old, metal-rich stars, the cluster value for [Ni/Fe] is above 0.00 by
0.17 $\pm$0.06 dex.

In disk and halo field stars the alpha-elements, Mg, Si, Ca, Ti show
enhancements in stars with lower metallicities (e.g. Edvardsson et al.~1993,
Stephens \& Boesgaard 2002), but these four elements usually share similar
enhancements.  There is, however, generally more scatter in the [Ti/Fe] values
at a given [Fe/H], for example see Figure 18 in Stephens \& Boesgaard which
shows the four alpha elements and includes the results of Edvardsson et
al.~(1993).  There is a range of $\sim$0.3 dex in [Ti/Fe] for field stars --
of all ages -- in the Edvardsson et al.~(1993) sample with [Fe/H] $>$0.0.  The
mild enhancement in [Ti/Fe] that may be present in the NGC 6791 stars could
result from a combination of our errors and an intrinsic scatter in this
alpha-product element.

The apparent Ni enhancement could be the result of the contribution of SN Ia
to the pre-cluster gas.  The models of exploding CO white dwarfs of Tsujimoto
et al.~(1995) predict that the mass of Ni/Fe that is synthesized could be a
factor of two higher than the solar ratio.

As we point out in $\S$ 2, NGC 6791 does not fit any age-metallicity relation
in the Galactic disk.  Nor do the old, metal-rich field stars field stars we
have used for our comparison sample in Figure 4 fit the age-metallicity
relationship.  Furthermore, as Carraro et al.~(2006) point out, NGC 6791 does
not fit a radial abundance gradient in the Galaxy.  The existence of such a
cluster indicates that there were complex beginnings and subsequent evolution
of our Galaxy.  The orbital solution for the cluster of Bedin et al.~(2006)
indicate that it could have been formed in the inner regions of the Galaxy.
Inasmuch as the Galactic bulge is metal-rich -- presumably due to rapid early
star formation and chemical enrichment -- the formation of NGC 6791 in that
region could account for the enhancement of our element abundances.  Only
because of its high cluster mass and large stellar density has NGC 6791
survived through several crossings of the Galactic disk.

\section{SUMMARY AND CONCLUSIONS}

NGC 6791 is an exotic open cluster in part because its metallicity is at least
a factor of two higher than solar, yet it is very old.  Other old open
clusters such as M67 and NGC 188 with ages of 4-7 Gyr are not metal-rich but
seem to have solar or near-solar composition.  With its mass of $\sim$4000
M$_{\odot}$, NGC 6791 is especially massive for an open cluster.

Previous determinations of the metallicity of NGC 6791 have been of evolved
stars because the main sequence stars are faint and the available telescopes
of 4-5 m in diameter.  We have been able to obtain high-resolution spectra of
two unevolved turnoff stars with the Keck I 10 m telescope with HIRES.

Due to the great distance and low Galactic latitude of NGC 6791 and the
consequent large (and uncertain) value for the reddening, we relied on our
spectra for the stellar parameter determination.  In addition to finding the
abundance of Fe, we have determined abundances for Na, Si, Ca, Ti, Cr, Ni, Y
and Ba.

We have compared our abundance results for our turnoff stars with a sample of
19 old (age $>$8 Gyr), metal-rich ([Fe/H] $>$+0.01) field stars from the
literature.  The abundance ratios in the cluster and those field stars are
very similar, with the probable exception of [Ni/Fe].  An enhancement in Ni
could result from a surfeit of SN Ia explosions in the pre-cluster gas.
Grenon (1999) has studied the origin and kinematics of super-metal-rich stars
([M/H] $>$ +0.30) that are 10 Gyr old and suggests that they formed close to
the bulge and, through perturbations by the bar of the Galaxy, were driven to
greater distances from the center.  This could be the case for our cluster and
for the 19 field comparison stars.  The orbits calculated by Bedin et
al.~(2006) indicate that NGC 6791 has a perigalacticon of $\sim$3 kpc, an
apogalacticon of $\sim$10 kpc, and a {\it boxy} type orbit with a high
eccentricity, $\sim$0.5: unusual for an open cluster.

In any case, it seems clear that formation and history of the Galaxy and the
Galactic disk were not simple, straightforward processes.  The beginnings for
galactic structure and chemical evolution must have been complex.  NGC 6791 is
an example of this complexity: massive for an open cluster, old yet
super-metal-rich, an eccentric orbit, currently a kpc above the Galactic
plane, yet having undergone several passages through the disk, inconsistent
with any radial abundance gradient and in defiance of any age-metallicity
relation.

\acknowledgements We are grateful to Aaron Steinhauer and Steven Margheim for
help with the data collection and to Elizabeth McGrath and Hai Fu for help
with the data reduction.  This work was supported by NSF AST 0097955 and NSF
AST 0505899 to AMB and to the REU site grant which supported EECJ at the
University of Hawaii, NSF AST 0453395, and to NSF AST 9812735 to CPD.

\begin{figure}
\plotone{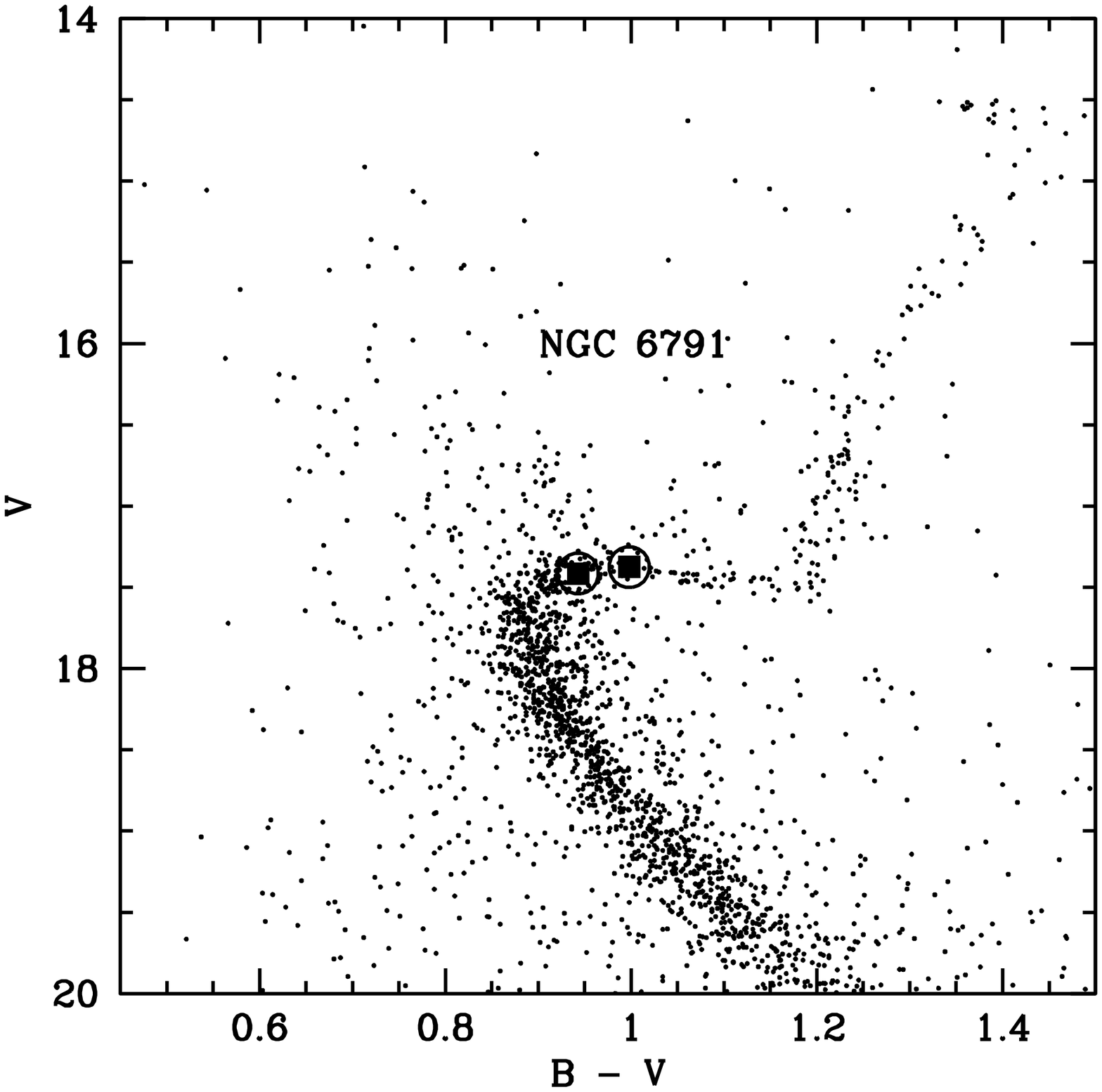}
\caption{The color-magnitude diagram of NGC 6791 from the photometry of MJP.
The two turnoff stars that we observed are indicated by the large circles.}
\end{figure}

\begin{figure}
\plotone{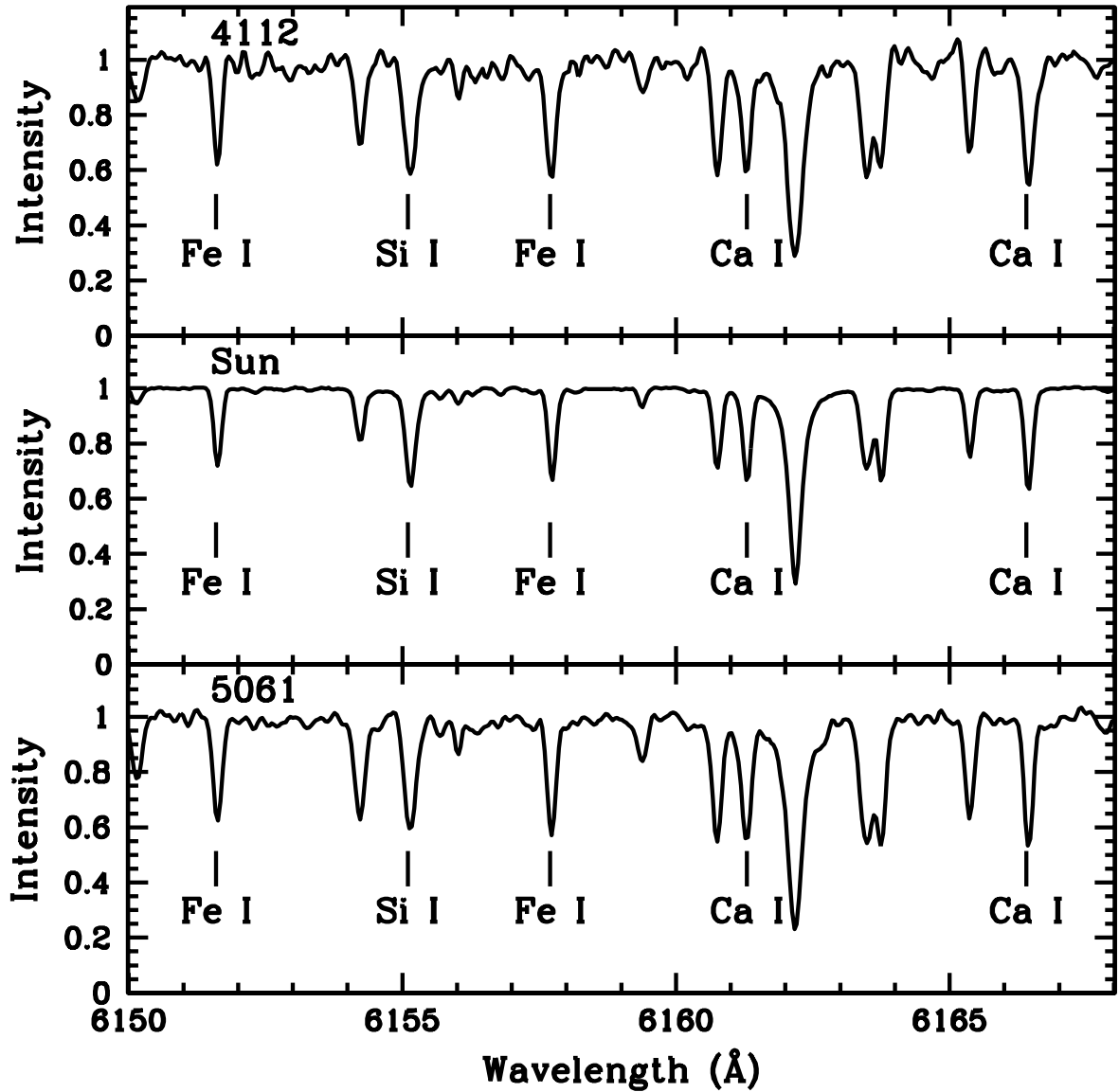}
\caption{Sample spectral regions of each star along with a 10 s exposure of
the spectrum of the Moon taken with HIRES on 2003 Jan. 11.  Lines of Si I, Ca
I and Fe I are indicated.  MJP 4112 is hotter than the Sun at 5800 K while MJP
5061 is cooler at 5650 K; notice that the lines are stronger in both the
hotter and the cooler star in NGC 6791 relative to solar.}
\end{figure}

\begin{figure}
\plotone{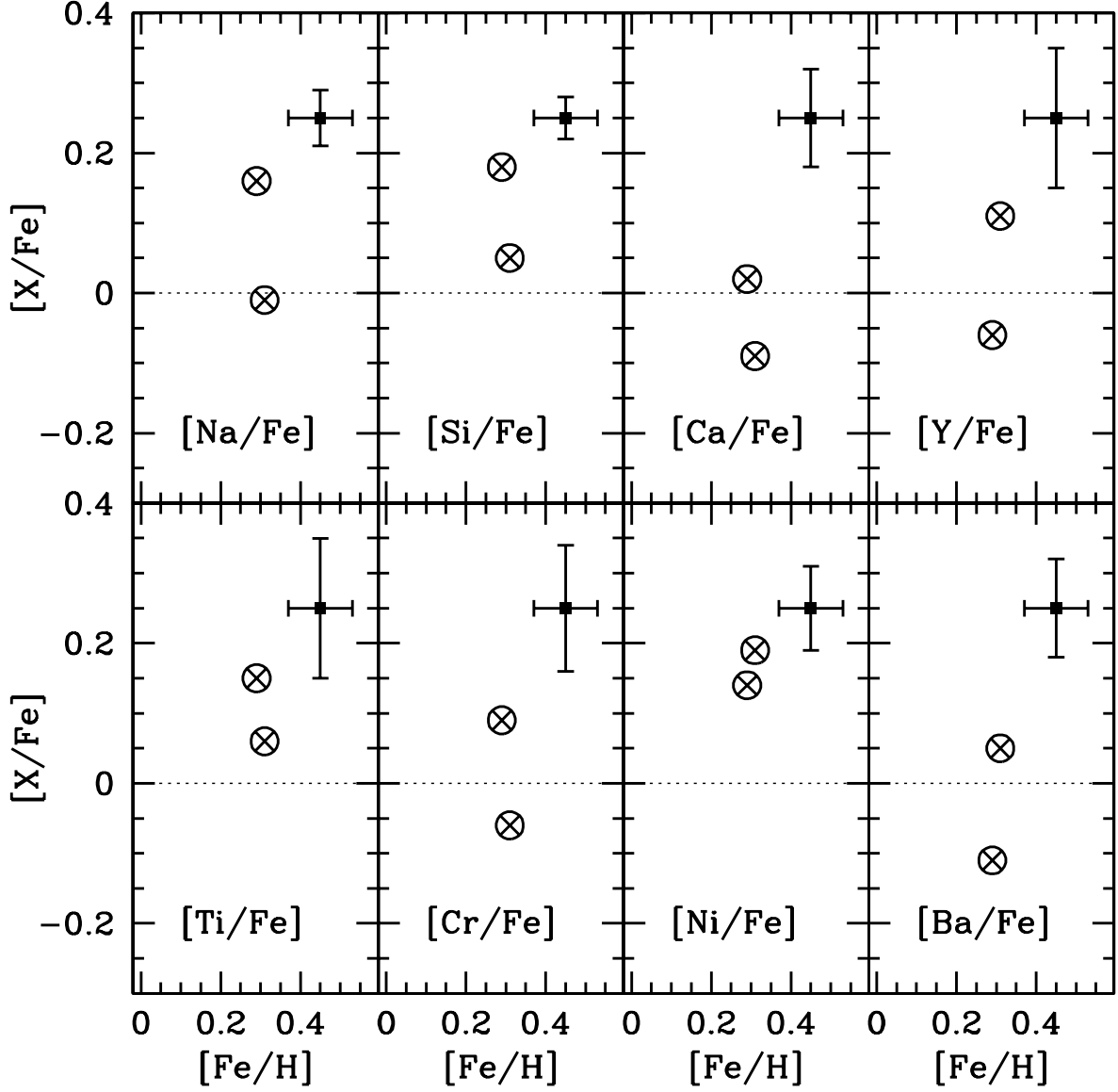}
\caption{Abundance ratios relative to Fe in the two stars of NGC 6791.  The
1$\sigma$ errors due to uncertainties in the stellar parameters for the
cluster mean are shown for each element ratio.  Within the errors the ratios
[X/Fe] are $\sim$0.0 for the cluster mean with the probable exception of Ni.}
\end{figure}

\begin{figure}
\plotone{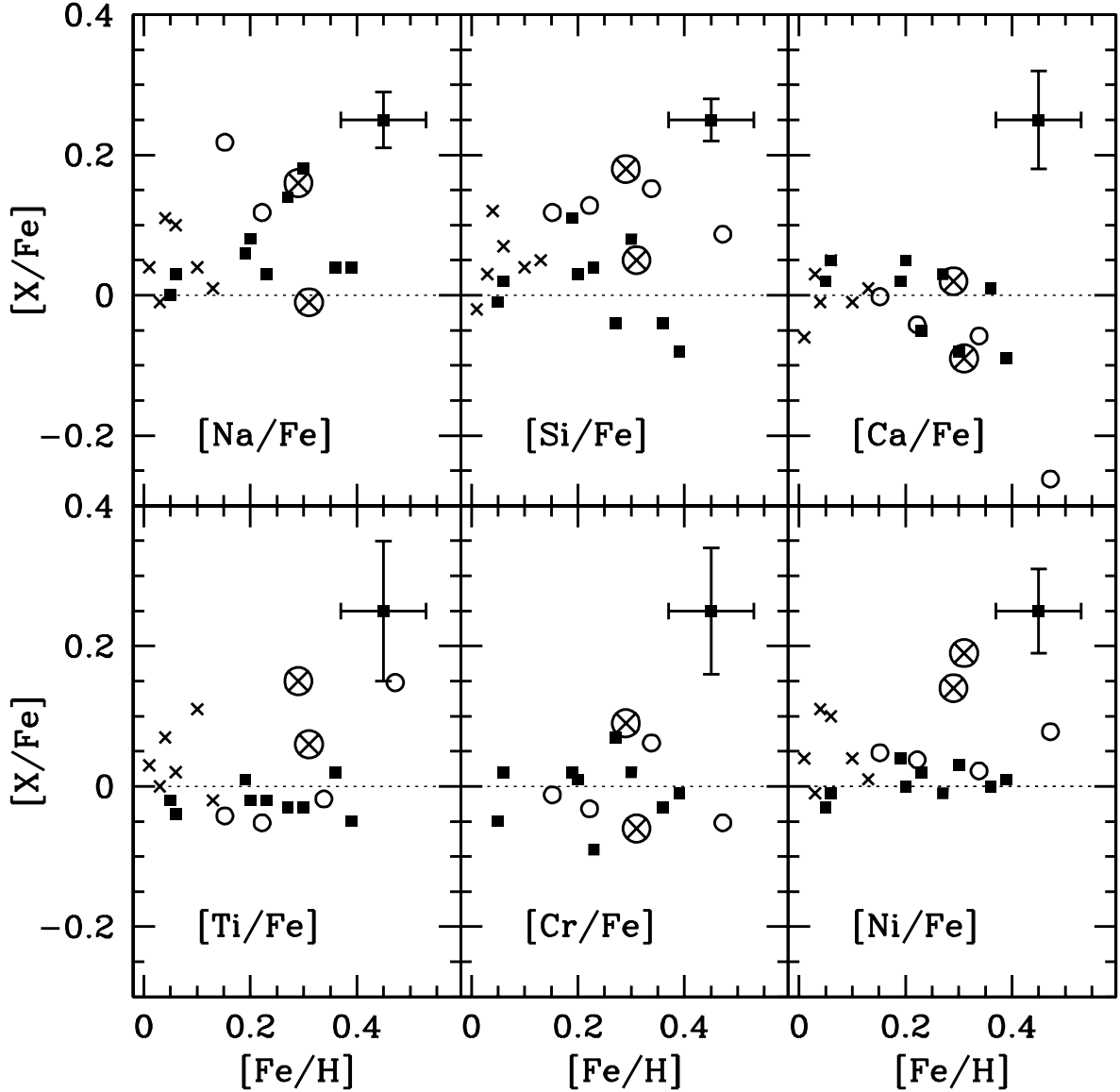}
\caption{Comparison of the abundance results in our turnoff stars in NGC 6791
with old, metal-rich field stars.  Our data are the large circled crosses.
The small crosses are from Edvardsson et al.~(1993), the open circles are from
Feltzing \& Gonzales (2001). and the filled squares are from Chen et
al.~(2003).  Those field stars are all older than 8 Gyr.  The two cluster
stars fall in the same region as the field stars, with the probable exception
of [Ni/Fe] which is nearly 3$\sigma$ above 0.0.  Certainly the cluster mean
values from Table 7 are completely consistent with the field stars, except for
Ni.  The 1$\sigma$ errors due to uncertainties in the stellar parameters are
shown for each element ratio and refer only to our stars.}
\end{figure}

\clearpage

\singlespace
\begin{center}
\begin{deluxetable}{lrcrccc} 
\tablewidth{0pc}
\tablecolumns{7} 
\tablecaption{Keck HIRES Observations} 
\tablehead{ 
\colhead{Star}  &  \colhead{V}  & \colhead{B-V} &
\colhead{Night} & \colhead{Exp. Time} & \colhead {S/N} & \colhead{Total}  \\

\colhead{} & \colhead{} & \colhead{} & \colhead{}  & \colhead{(min)} &
\colhead{} & \colhead{S/N}} 
\startdata 
4112 &  17.415   & 0.943 &  28 May 2000  & 120 & 30 & \phn \\
\phn &  \phn     & \phn  &  29 May 2000  & 120 & 24 & 38 \\
5061 &  17.375   & 0.998 &  07 Jun 1999  & \phn 80 & 25 & \phn \\
\phn &  \phn     & \phn  &  08 Jun 1999  & 100 & 31 & 40 \\
\enddata 

\end{deluxetable} 
\end{center}



\singlespace
\begin{center}
\begin{deluxetable}{lcccc} 
\tablewidth{0pc}
\tablenum{3}
\tablecolumns{5}
\tablecaption{Model Parameters} 
\tablehead{ 
\colhead{Star} & \colhead{T$_{\rm eff}$ (K)} & \colhead{log g} 
& \colhead{[Fe/H]} & \colhead{$\xi$} (km s$^{-1})$}

\startdata 
MJP 4112 & 5800 & 4.3 & +0.3 & 1.35 \\
MJP 5061 & 5650 & 4.3 & +0.3 & 1.23 \\
\enddata 

\end{deluxetable} 
\end{center}


\clearpage

\singlespace
\begin{center}
\begin{deluxetable}{lccccclccc} 
\tablenum{4}
\tablewidth{0pc}
\tablecolumns{10} 
\tablecaption{Abundances Results} 
\tablehead{ 

\colhead{MJP 4112} & \colhead{} & & & & & \colhead{MJP 5061} & & & \\

\colhead{Ion}  & \colhead{Abund.} & \colhead{\# lines} &
\colhead{st.dev} & & & \colhead{Ion}  & \colhead{Abund.} 
& \colhead{\#lines} & \colhead{st.dev}} 

\startdata 
Na I	&  6.78  & 2  & $\pm$0.03 & & &  Na I & 6.63 & 2  & $\pm$0.10 \\
Si I	&  8.02	 & 7  & $\pm$0.17 & & & Si I & 7.91 & 6  & $\pm$0.13 \\ 
Ca I	&  6.67	 & 19 & $\pm$0.12 & & & Ca I  &  6.59 & 23 & $\pm$0.16 \\ 
Ti I	&  5.43	 & 15 & $\pm$0.21 & & & Ti I &  5.36 & 15 & $\pm$0.17 \\
Cr I	&  6.00	 & 15 & $\pm$0.16 & & & Cr I &  5.91 & 14 & $\pm$0.18 \\ 
Cr II	&  6.29	 & 3  & $\pm$0.13 & & & Cr II &  5.96 & 4  & $\pm$0.07 \\
Cr	&  6.05 &  18 & \nodata & & &   Cr    &  5.92 & 18 & \nodata  \\
Fe I	&  7.78	 & 32 & $\pm$0.14 & & & Fe I &  7.84 & 35 & $\pm$0.13 \\
Fe II	&  7.97	 & 5  & $\pm$0.06 & & & Fe II &  7.76 & 4  & $\pm$0.03 \\
Fe	&  7.81  & 37 & \nodata & & & Fe & 7.83 & 39 & \nodata \\
Ni I	&  6.68	 & 19 & $\pm$0.15 & & & Ni I &  6.75 & 24 & $\pm$0.14 \\ 
Y II	&  2.47  & 3  & $\pm$0.34 & & & Y II &  2.66 & 3  & $\pm$0.14 \\
Ba II   &  2.31  & 3  & $\pm$0.09 & & & Ba II & 2.49 & 2 &  $\pm$0.25 \\
\enddata 

\end{deluxetable} 
\end{center}



\singlespace
\begin{center}
\begin{deluxetable}{lcccc} 
\tablenum{5}
\tablewidth{0pc}
\tablecolumns{5} 
\tablecaption{Abundance Errors for MJP 4112} 
\tablehead{ 

\colhead{Ion} & \colhead{T=$\pm$100K} & \colhead{log g=$\pm$0.2} &
\colhead{[Fe/H]=$\pm$0.1} & \colhead{$\xi$=$\pm$0.2}  
}
\startdata 
Na I    & $\pm$0.05   & $\mp$0.03  & $\pm$0.00 & $\mp$0.02 \\
Si I    & $\pm$0.02   & $\mp$0.02  & $\pm$0.01 & $\mp$0.02 \\
Ca I    & $\pm$0.07   & $\mp$0.06  & $\pm$0.02 & $\mp$0.04 \\
Ti I    & $\pm$0.10   & $\mp$0.03  & $\pm$0.00 & $\mp$0.07 \\
Cr I    & $\pm$0.10   & $\mp$0.05  & $\pm$0.01 & $\mp$0.07 \\
Cr II   & $\mp$0.04   & $\pm$0.03  & $\pm$0.03 & $\mp$0.08 \\
Fe I    & $\pm$0.08   & $\mp$0.03  & $\pm$0.01 & $\mp$0.07 \\
Fe II   & $\mp$0.07   & $\pm$0.03  & $\pm$0.06 & $\mp$0.04 \\
Ni I    & $\pm$0.06   & $\mp$0.02  & $\pm$0.01 & $\mp$0.05 \\
Y II    & $\pm$0.00   & $\pm$0.05  & $\pm$0.04 & $\mp$0.10 \\
Ba II   & $\pm$0.01   & $\pm$0.01  & $\pm$0.04 & $\mp$0.08 \\
\enddata 

\end{deluxetable} 
\end{center}


\clearpage

\singlespace
\begin{center}
\begin{deluxetable}{lcccc} 
\tablenum{6}
\tablewidth{0pc}
\tablecolumns{5} 
\tablecaption{Abundance Errors for MJP 5061} 
\tablehead{ 

\colhead{Ion} & \colhead{T=$\pm$100K} & \colhead{log g=$\pm$0.2} &
\colhead{[Fe/H]=$\pm$0.1}  & \colhead{$\xi$=$\pm$0.2}
}
\startdata 
Na I    & $\pm$0.05   & $\mp$0.03  & $\pm$0.01 & $\mp$0.01 \\
Si I    & $\pm$0.00   & $\mp$0.01  & $\pm$0.02 & $\mp$0.01 \\
Ca I    & $\pm$0.07   & $\mp$0.06  & $\pm$0.02 & $\mp$0.03 \\
Ti I    & $\pm$0.11   & $\mp$0.03  & $\pm$0.01 & $\mp$0.05 \\
Cr I    & $\pm$0.10   & $\mp$0.04  & $\pm$0.02 & $\mp$0.05 \\
Cr II   & $\mp$0.04   & $\pm$0.04  & $\pm$0.03 & $\mp$0.04 \\
Fe I    & $\pm$0.07   & $\mp$0.03  & $\pm$0.02 & $\mp$0.04 \\
Fe II   & $\mp$0.06   & $\pm$0.06  & $\pm$0.04 & $\mp$0.03 \\
Ni I    & $\pm$0.05   & $\mp$0.02  & $\pm$0.02 & $\mp$0.03 \\
Y II    & $\pm$0.00   & $\pm$0.06  & $\pm$0.04 & $\mp$0.07 \\
Ba II   & $\pm$0.02   & $\pm$0.01  & $\pm$0.06 & $\mp$0.08 \\
\enddata 

\end{deluxetable} 
\end{center}



\singlespace
\begin{center}
\begin{deluxetable}{lcccclcccccc} 
\tablenum{7}
\tablewidth{0pc}
\tablecolumns{12} 
\tablecaption{Abundances Ratios and Errors} 
\tablehead{ 

\colhead{MJP 4112} & \colhead{} & & & & \colhead{MJP 5061} & & & & & \colhead{Cluster}
& \\

\colhead{Element}  & \colhead{Abund.} & \colhead{$\sigma$} & & &
   \colhead{Element}  & \colhead{Abund.} 
& \colhead{$\sigma$} & & & \colhead{Mean}  & \colhead{$\sigma$}  }

\startdata
${\rm [Fe/H]}$	& +0.29 & $\pm$0.11 & & & [Fe/H]  & +0.31 &   $\pm$0.11 & & 
& +0.30 & $\pm$0.08 \\
${\rm [Na/Fe]}$ & +0.16 & $\pm$0.06 & & & [Na/Fe] & $-$0.01 & $\pm$0.06 & & 
& +0.07 & $\pm$0.04 \\
${\rm [Si/Fe]}$	& +0.18 & $\pm$0.04 & & & [Si/Fe] & +0.05 &   $\pm$0.03 & & 
& +0.11 & $\pm$0.03 \\
${\rm [Ca/Fe]}$	&  0.02 & $\pm$0.10 & & & [Ca/Fe] & $-$0.09 & $\pm$0.10 & & 
& $-$0.03 & $\pm$0.07 \\
${\rm [Ti/Fe]}$	& +0.15 & $\pm$0.13 & & & [Ti/Fe] & +0.06 &   $\pm$0.14 & & 
& +0.10 & $\pm$0.10 \\
${\rm [Cr/Fe]}$	& +0.09 & $\pm$0.12 & & & [Cr/Fe] & $-$0.06 & $\pm$0.13 & & 
& +0.01 & $\pm$0.09 \\
${\rm [Ni/Fe]}$	& +0.14 & $\pm$0.08 & & & [Ni/Fe] & +0.19 &   $\pm$0.08 & & 
& +0.17 & $\pm$0.06 \\
${\rm [Y/Fe]}$& $-$0.06 & $\pm$0.12 & & & [Y/Fe] & +0.11 &    $\pm$0.14 & & 
& +0.03 & $\pm$0.10 \\  
${\rm [Ba/Fe]}$& $-$0.11 & $\pm$0.09 & & & [Ba/Fe] & +0.05 &    $\pm$0.10 & & 
& $-$0.03 & $\pm$0.07 \\  
\enddata 

\end{deluxetable} 
\end{center}


\appendix

\tightenlines


\begin{deluxetable} {lrrrr}
\tablecolumns{5}
\tablewidth{0pc}
\tablenum{2}
\tablecaption{Measured Equivalent Widths}
\tablehead{
\multicolumn{1}{c}{$\lambda$ } &
\multicolumn{1}{c}{Ex. Pot.} &
\multicolumn{1}{c}{$\log{gf}$} &
\multicolumn{2}{c}{Equivalent Widths}\\
\multicolumn{1}{c}{($\rm{\AA}$)} &
\multicolumn{1}{c}{(eV)} &
\multicolumn{1}{c}{} &
\multicolumn{1}{c}{MJP 4112} &
\multicolumn{1}{c}{MJP 5061} 
}
\startdata
\multicolumn{1}{l}{\ion{Na}{1}} \\
\hline
6154.230 & 2.10 & $-$1.66 & 69.7 & 55.8 \\
6160.753 & 2.10 & $-$1.35 & 93.8 & 96.0 \\
\hline
\multicolumn{1}{l}{\ion{Si}{1}} \\
\hline
5772.148 & 5.08  & $-$1.75 &  90.8 & \nodata  \\
5948.545 & 5.08  & $-$1.225 & 121.6 & 110.3  \\
6125.026 & 5.61  & $-$1.48  &  52.5 & 70.3 \\
6142.490 & 5.62  & $-$1.48  &  77.6 &  54.5  \\
6145.020  & 5.61 & $-$1.37 &  59.1 &  55.5  \\
6155.141  & 5.62 & $-$0.84 & 121.0 & 112.0  \\
6243.820  & 5.61 & $-$1.27 & 105.7 &  85.5  \\

\hline
\multicolumn{1}{l}{\ion{Ca}{1}} \\
\hline
4512.270  & 2.52  & $-$1.90 &  37.6 &  57.6 \\
4526.934 & 2.71  & $-$0.49 & 146.1 & 126.9 \\
4685.268 & 2.93  & $-$1.88 &  95.5 &  85.2 \\
5260.389 & 2.52  & $-$1.72 &  67.3 &  69.1 \\
5261.707 & 2.52  & $-$0.6545 & 121.7 & 129.1 \\
5262.241 & 2.52  & $-$0.60 & 141.5 & 161.6 \\
5512.980 & 2.93  & $-$0.3685 & 134.6 & 119.7 \\
5581.968 & 2.52  & $-$0.56 & \nodata & 131.6 \\
5590.117 & 2.52  & $-$0.6405 & 127.9 & 113.8 \\
5594.466 & 2.52  &  0.023    & 189.6 & \nodata \\
5598.480 & 2.52  & $-$0.22 & \nodata & 172.5 \\ 
5857.451 & 2.93  & 0.235  & \nodata & 171.1 \\
6122.217 & 1.89  & $-$0.32 & 232.6 & 246.5 \\
6161.300 & 2.52  & $-$1.27 &  99.7 & 104.2 \\
6162.173 & 1.90  & $-$0.09 & 264.2 & 261.5 \\
6163.755 & 2.52  & $-$1.286 & \nodata & 100.0 \\
6166.440 & 2.52  & $-$1.14 & \nodata &  97.1 \\
6169.042 & 2.52  & $-$0.797 & 132.6 & 116.1 \\
6169.562 & 2.53  & $-$0.374 & 157.0 & 145.8 \\
6449.810 & 2.52  & $-$0.502 & 142.0 & 146.9 \\
6455.600 & 2.52  & $-$1.34 &  97.5 &  96.2 \\
6462.569 & 2.52  &  0.286    & 288.3 & 285.4 \\
6471.688 & 2.52  & $-$0.690  & 132.2 & 135.7 \\
6499.650 & 2.52  & $-$0.818 & 120.1 & 124.5 \\
\hline
\tablebreak
\multicolumn{1}{l}{\ion{Ti}{1}} \\
\hline
4518.023 & 0.826  & $-$0.269 & 119.8 & \nodata \\
4533.239 & 0.848  &  0.532 & 178.1 & 150.6 \\
4534.778 & 0.836  &  0.336 & \nodata & 144.8 \\
4617.254 & 1.749  &  0.445 & 85.7 & \nodata \\
4840.874 & 0.900  & $-$0.453 & 82.8 & 102.5 \\
4981.732 & 0.848  &  0.560 & 153.1 & \nodata \\
4991.067 & 0.836  &  0.436 & \nodata & 155.3 \\
4999.504 & 0.826  &  0.306 & 146.7 & 140.2 \\
5016.162 & 0.848  & $-$0.518 & 108.8 & 98.8 \\
5020.028 & 0.836  & $-$0.358 & \nodata & 118.2 \\
5022.871 & 0.826  & $-$0.378 & 111.6 & 110.9 \\
5036.468 & 1.443  &  0.186 & \nodata & 95.3 \\
5039.959 & 0.021  & $-$1.130 & \nodata & 122.8 \\
5064.654 & 0.048  & $-$0.855 & \nodata & 136.9 \\
5192.969 & 0.021  & $-$0.950 & 107.3 & 126.6 \\
5953.170 & 1.890  & $-$0.310  & 74.5 & 77.1 \\
5978.540 & 1.870  & $-$0.440 & 71.2 & \nodata \\
6126.224 & 1.070  & $-$1.320  & 56.2 & 61.6 \\
6258.110 & 1.440  & $-$0.430  & 72.2 & 93.1 \\
6261.106 & 1.430  & $-$0.480  & 77.3 & \nodata \\
6303.760 & 1.440  & $-$1.600  & 27.0 & \nodata \\
\hline
\multicolumn{1}{l}{\ion{Cr}{1}} \\
\hline
4511.900 & 3.0900  & $-$1.150 & 65.8 & 71.5 \\
4545.945 & 0.9415  & $-$1.370 & 100.2 & \nodata \\
4591.389 & 0.9685  & $-$1.740 & \nodata & 89.6 \\ 
4600.741 & 1.0037  & $-$1.260 & 130.9 & 101.4 \\
4616.120 & 0.9829  & $-$1.190 & 112.9 & 134.5 \\
4626.174 & 0.9685  & $-$1.320 & 115.4 & 107.4 \\
4651.282 & 0.9829  & $-$1.460 & 117.3 & 103.6 \\
4652.152 & 1.0037  & $-$1.030 & 124.8 & 140.7 \\  
4789.350 & 2.5400  & $-$0.366 & 80.7  & 80.3 \\
5206.038 & 0.9415  &  0.019 & 385.4 & \nodata \\
5247.566 & 0.9610  & $-$1.640 & 107.9 & 122.5 \\
5296.691 & 0.9829  & $-$1.400 & 128.9 & 124.5 \\
5298.277 & 0.9829  & $-$1.160 & \nodata & 162.3 \\ 
5329.170 & 2.9100  & $-$0.064 & 98.6 & \nodata \\
5345.801 & 1.0037  & $-$0.980 & 178.6 & 175.0 \\ 
5348.312 & 1.0037  & $-$1.290 & 142.0 & 136.1 \\
5409.772 & 1.0301  & $-$0.715 & 192.2 & \nodata \\
6330.100 & 0.940   & $-$2.94   & \nodata & 58.2 \\
\hline
\multicolumn{1}{l}{\ion{Cr}{2}} \\
\hline
4558.650 & 4.0737  & $-$0.660 & \nodata & 85.9 \\ 
4588.199 & 4.0715  & $-$0.630 & \nodata & 87.9 \\
4634.070 & 4.0725  & $-$1.240 & 82.0 & \nodata \\ 
4848.235 & 3.8647  & $-$1.140 & 87.6 & 79.3 \\ 
5237.329 & 4.0737  & $-$1.160 & 90.5 & 62.3 \\
\hline
\multicolumn{1}{l}{\ion{Fe}{1}} \\
\hline
5916.247 & 2.450   & $-$2.9900 &  \nodata &  90.6 \\
5934.655 & 3.930   & $-$1.0200 & 114.8 & 106.5 \\
5956.700 & 0.860   & $-$4.5640 &  78.9 &  86.1 \\
6024.058 & 4.550   & $-$0.0600 & 135.2 & 139.0 \\
6027.048 & 4.076   & $-$1.1495 & 103.4 &  88.6 \\
6055.992 & 4.734   & $-$0.4600 &  99.4 &  93.9 \\
6065.481 & 2.609   & $-$1.4700 & 156.8 & 160.0 \\
6078.999 & 4.652   & $-$1.1200 & \nodata & 75.5 \\
6082.720 & 2.220   & $-$3.5330 & 65.5  & 68.7 \\
6127.904 & 4.143   & $-$1.3990 &  63.2 & 75.0 \\
6136.615 & 2.453   & $-$1.4050 & 171.8 & \nodata \\ 
6137.691 & 2.588   & $-$1.3745 & 181.8 & 193.5 \\
6151.618 & 2.176   & $-$3.2990 &  71.4 & 75.4 \\
6157.725 & 4.076   & $-$1.2600 & \nodata & 87.1 \\
6165.360 & 4.140   & $-$1.4700 &  66.1 & 73.5 \\
6173.341 & 2.223   & $-$2.8800 &  96.3 & 102.2 \\
6180.203 & 2.728   & $-$2.6225 &  81.9 &  94.8 \\
6187.990 & 3.940   & $-$1.5700 &  74.5 &  81.7 \\
6219.280 & 2.198   & $-$2.4330 & 110.5 & \nodata \\
6230.723 & 2.559   & $-$1.2785 & 183.6 & 230.1 \\
6246.318 & 3.603   & $-$0.8770 & 156.4 & 162.9 \\
6252.555 & 2.404   & $-$1.7270 & 150.0 & 165.7 \\
6265.140 & 2.180   & $-$2.5100 & 107.9 & 126.4 \\
6297.800 & 2.220   & $-$2.7000 &  97.1 & 118.1 \\
6335.330 & 2.198   & $-$2.2035 & 119.1 & 133.1 \\
6355.029 & 2.85    & $-$2.358  &  82.6 & \nodata \\
6380.742 & 4.187   & $-$1.3875 &  86.5 &  82.7 \\
6393.601 & 2.433   & $-$1.5760 & 171.8 & 176.0 \\
6411.649 & 3.650   & $-$0.6600 & 149.7 & 177.4 \\
6469.193 & 4.830   & $-$0.6200 &  98.3 &  90.1 \\
6481.869 & 2.279   & $-$2.9840 &  90.7 & 105.7 \\
6498.940 & 0.958   & $-$4.6946 & \nodata & 86.5 \\ 
6592.913 & 2.728   & $-$1.5365 & \nodata & 153.8 \\
6593.868 & 2.433   & $-$2.3940 & \nodata & 125.7 \\
6597.561 & 4.795   & $-$0.9200 &  65.3 &  75.6 \\
6609.109 & 2.559   & $-$2.6765 &  97.1 & 102.1 \\
6750.152 & 2.424   & $-$2.6080 &  93.4 & 104.2 \\
6752.707 & 4.640   & $-$1.2000 &  57.9 &  69.1 \\
\hline
\multicolumn{1}{l}{\ion{Fe}{2}} \\
\hline
6084.100 & 3.20  & $-$3.810 & 39.9 & \nodata \\
6113.322 & 3.22  & $-$4.160 & 27.2 & \nodata \\   
6149.249 & 3.89  & $-$2.9300 & \nodata & 44.8 \\  
6247.562 & 3.89  & $-$2.7240 & 80.3 & 62.1 \\
6369.460 & 2.89  & $-$4.16   & 38.5 & 26.2 \\
6456.391 & 3.90  & $-$2.3290 & 90.4 & 79.0 \\
\hline
\multicolumn{1}{l}{\ion{Ni}{1}} \\
\hline
4714.408 & 3.3801  &  0.2300 & \nodata & 229.5 \\
4715.757 & 3.5435  & $-$0.3400 & 108.0 & 143.4 \\ 
4756.510 & 3.4802  & $-$0.3400 & 106.3 & \nodata \\
4786.531 & 3.4198  & $-$0.1700 & \nodata & 131.9 \\ 
4831.169 & 3.6063  & $-$0.4200 & 86.2 & 97.2 \\  
4904.407 & 3.5424  & $-$0.1700 & 129.7 & 155.6 \\
4937.341 & 3.6063  & $-$0.3900 & \nodata & 129.1 \\ 
5115.389 & 3.8342  & $-$0.1100 & 100.2 & \nodata \\ 
5155.762 & 3.8985  & $-$0.0900 & 129.5 & 100.9 \\
5476.900 & 1.8263  & $-$0.8900 & 196.3 & 217.6 \\
6086.288 & 4.26    & $-$0.530  &  79.0 & 73.3 \\
6108.125 & 1.68    & $-$2.60   & \nodata & 99.5 \\
6111.078 & 4.09    & $-$0.810  &  57.9 & 70.4 \\
6128.984 & 1.68    & $-$3.330  &  \nodata & 70.5 \\
6130.141 & 4.26    & $-$0.960  &  52.2 & 45.6 \\
6133.963 & 4.09    & $-$1.810  &  14.6 & \nodata \\
6175.360 & 4.09    & $-$0.560  &  73.5 & 70.2 \\
6176.816 & 4.09    & $-$0.260  &  98.2 & 98.3 \\
6177.236 & 1.83    & $-$3.560  &  30.0 & 43.6 \\
6186.709 & 4.11    & $-$0.920  & \nodata & 63.2 \\
6327.604 & 1.68    & $-$3.110  & \nodata & 73.2 \\
6378.260 & 4.15    & $-$0.850  &  68.7 & 54.4 \\
6482.809 & 1.93    & $-$2.830  &  65.0 & 80.1 \\
6586.319 & 1.95    & $-$2.730  &  76.4 & 74.6 \\
6635.130 & 4.42    & $-$0.740  &  38.4 & 58.0 \\
6767.768 & 1.8263  & $-$2.1700 & 116.2 & 123.9 \\ 
6772.321 & 3.66    & $-$0.950  & \nodata & 78.3 \\
\hline
\multicolumn{1}{l}{\ion{Y}{2}} \\
\hline
4883.684 & 1.0821  & $-$0.0100 & 74.1 & 73.0 \\ 
4900.120 & 1.0313  & $-$0.1300 & 84.6 & 82.6 \\ 
5087.416 & 1.0810  & $-$0.3100 & 43.6 & 65.0 \\ 
\hline
\multicolumn{1}{l}{\ion{Ba}{2}} \\
\hline
4554.000 &  0.000  &  0.17      & 185.5 & 194.8 \\ 
5853.700 &  0.000  & $-$1.0000 & 67.0 & \nodata \\
6141.700 &  0.000  & $-$0.0760 & 123.0 & 144.8 \\

\enddata
\end{deluxetable}

\end{document}